\documentclass[traditabstract]{aa}

\usepackage{graphicx}
\pdfoutput=1
\usepackage{amsmath}
\usepackage{txfonts}
\usepackage{natbib}
\usepackage[usenames,dvipsnames]{color}
\usepackage{enumerate}
\newcommand{\logg}{\log_{10} g}
\newcommand{\loggunits}{\log_{10}(g/\mathrm{cm}\,\mathrm{s}^{-2})}
\newcommand{\Teff}{T_{\mathrm{eff}}}

\newcommand{\cm}{\mathrm{cm}}
\newcommand{\mbarn}{\mathrm{mbarn}}
\newcommand{\Msun}{M_{\odot}}
\newcommand{\Rsun}{R_{\odot}}

\newcommand{\Hy}{\mathrm{H}}
\newcommand{\C}{\mathrm{C}}

\newcommand{\Fe}{\mathrm{Fe}}

\newcommand{\Ba}{\mathrm{Ba}}

\newcommand{\Eu}{\mathrm{Eu}}
\newcommand{\Pb}{\mathrm{Pb}}
\newcommand{\Bi}{\mathrm{Bi}}
\newcommand{\X}{\mathrm{X}}
\newcommand{\Y}{\mathrm{Y}}
\newcommand{\ls}{\mathrm{ls}}
\newcommand{\hs}{\mathrm{hs}}
\begin{document}


   \title{How plausible are the proposed\\ formation scenarios of CEMP-r/s stars?}

   \author{Carlo Abate
          \and
          Richard J. Stancliffe
          \and
          Zheng-Wei Liu
          }

   \institute{Argelander Institut f\"ur Astronomie, Auf dem H\"ugel 71, D-53121 Bonn, Germany\\\email{cabate@uni-bonn.de}
             }

   \date{Received ...; accepted ...}


\abstract
{
	{
	CEMP-$r/s$ stars are metal-poor stars with enhanced abundances of carbon and heavy elements associated with the \emph{slow} and \emph{rapid} neutron capture process ($s$- and $r$-elements, respectively). It is believed that carbon and $s$-elements were accreted in the past from the wind of a primary star in the asymptotic giant branch (AGB) phase of evolution, a scenario that is generally accepted to explain the formation of CEMP stars that are only enhanced in $s$-elements (CEMP-$s$ stars). The origin of $r$-element-enrichment in CEMP-$r/s$ stars is currently debated and many formation scenarios have been put forward.
 	}%
	{
	We aim to determine the likelihood of the scenarios proposed to explain the formation of CEMP-$r/s$ stars.
	}%
	{
	We calculate the frequency of CEMP-$r/s$ stars among CEMP-$s$ stars for a variety of formation scenarios, and we compare it with that determined from an observed sample of CEMP-$r/s$ stars collected from the literature.
	}%
	{
	The theoretical frequency of CEMP-$r/s$ stars predicted in most formation scenarios underestimates the observed ratio by at least a factor of five. If the enrichments in $s$- and $r$-elements are independent, the model ratio of CEMP-$r/s$ to CEMP-$s$ stars is about $22\%$, that is approximately consistent with the lowest estimate of the observed ratio. However, this model predicts that about one third of all carbon-normal stars have [Ba/Fe] and [Eu/Fe] higher than one, and that $40\%$ of all CEMP stars have $[\Ba/\Eu]\le0$. Stars with these properties are at least ten times rarer in our observed sample.
	}%
	{
	The \emph{intermediate} or $i$-process, which is supposedly active in some circumstances during the AGB phase, could provide an explanation of the origin of CEMP-$r/s$ stars, similar to that of CEMP-$s$ stars, in the context of wind mass accretion in binary systems. Further calculations of the nucleosynthesis of the $i$-process and of the detailed evolution of late AGB stars are needed to investigate if this scenario predicts a CEMP-$r/s$ star frequency consistent with the observations.
	}%
}
	\keywords{stars: abundances -- stars: AGB and post-AGB -- binaries: general -- stars: chemically peculiar -- stars: Population II -- Galaxy: halo}

\nopagebreak   \maketitle


\section{Introduction}
\label{intro}

Elements heavier than the iron group are mostly produced by neutron-capture processes. The canonical picture distinguishes two types of neutron-capture process: the \emph{slow} or $s$-process, in which the neutron densities are relatively low, $N_{\mathrm{n}}\approx10^7-10^8\,\cm^{-3}$; and the \emph{rapid} or $r$-process, in which the neutron densities are much higher, $N_{\mathrm{n}}\geq 10^{23}\,\cm^{-3}$ \cite[][]{Pagel2009}. Three parameters essentially determine what isotopes are produced by either neutron-capture process: the neutron exposure, that is the time-integrated neutron flux, the neutron-capture cross-sections of the isotopes, and the $\beta$-decay timescales of unstable isotopes. At low neutron exposures, $\tau_0 \lesssim 1\,\mbarn^{-1}$ \cite[e.g.][]{Kappeler1989, Busso1999, Pagel2009}, the neutron-capture timescale is comparable or longer to that of most $\beta$-decays, hence nucleosynthesis by the $s$-process progresses along the $\beta$-stability valley up to bismuth ($^{209}\Bi$). In contrast, at high neutron exposures the rate of neutron captures exceeds that of the $\beta$-decays. Consequently, very neutron-rich unstable nuclei are synthetised and, when the neutron flux is interrupted, they subsequently decay into stable isotopes on the neutron-rich side of the valley of stability \cite[][]{Seeger1965, Arnould2007, Pagel2009}. 

The $r$- and $s$-processes are active in completely different physical conditions. The latter is believed to take place in the intershell region of low-mass stars along the asymptotic giant branch \cite[AGB, e.g.][]{Truran1981-1, Gallino1998, Busso1999, Herwig2005} or in the helium-burning cores of massive stars \cite[][]{Peters1968, Truran1981-1, Pignatari2008, Pignatari2010, Frischknecht2012}. The site of $r$-nucleosynthesis is as yet uncertain, and it is normally associated with the collapse of massive-star cores into neutron stars, for example in supernova explosions and accretion-induced collapses \cite[][]{Burbidge1957, Cameron1957, Woosley1994, Qian2003, Cowan2004, Sneden2008}, or to the mergers of compact binary systems \cite[][]{RamirezRuiz2015}. 

Abundances of $r$- and $s$-elements are determined in stars of all metallicities and evolutionary stages to study the relative importance of the $r$- and $s$-processes at different epochs and investigate galactic chemical evolution \cite[e.g.][]{Spite1978, Truran1981-1, Burris2000, Travaglio2001, Mashonkina2003, Simmerer2004, Travaglio2004, Cescutti2006, Pagel2009, Roederer2013}. In the past three decades the abundances of neutron-capture elements have been determined in many studies which analyse high-resolution spectra of hundreds of stars (e.g. \citealp{Burris2000}, \citealp{Simmerer2004}, \citealp{Aoki2007, Aoki2013-1}, \citealp{Roederer2014-2}, and references therein). These studies show that high abundances of neutron-capture elements are very frequently found in carbon-enhanced metal-poor (CEMP) stars: i.e. in stars that are metal-poor (MP, i.e. with $[\Fe/\Hy]<-1$)%
\footnote{$[\X/\Y] = \log_{10}(N_{\X}/N_{\Y}) - \log_{10}(N_{\X}/N_{\Y})_{\odot}$, where $N_{\X}$ and $N_{\Y}$ are the number densities of elements X and Y, respectively.} 
stars with high carbon abundances, $[\C/\Fe]>1$. 

CEMP stars are observed as a substantial fraction of the MP stars in the halo, and the cumulative fraction of CEMP stars raises with decreasing metallicity, from three percent at $[\Fe/\Hy]<-1$ up to about $80\%$ at $[\Fe/\Hy]<-4$ \cite[e.g.][]{Cohen2005, Marsteller2005, Frebel2006, Lucatello2006, Lee2013, Yong2013III, Placco2014}. \cite{Aoki2007} show that at least $80\%$ of all CEMP stars in our Galaxy are also enriched in barium, whose solar-system abundance is mostly produced by the $s$-process \cite[$\approx 89\%$,][]{Bisterzo2011}, and are therefore called CEMP-$s$ stars. To explain the simultaneous enrichments in carbon, barium, and other $s$-elements it has been suggested that CEMP-$s$ stars underwent a mass-transfer episode from a more massive binary companion in the AGB phase of evolution. This scenario is supported by the evidence that most CEMP-$s$ stars show radial-velocity variations consistent with orbital motions \cite[][]{Lucatello2006, Starkenburg2014, Hansen2015-4}. 

Many CEMP stars are enriched in elements whose solar-system abundances are dominated by the $r$-process, such as europium \cite[$94\%$ of which comes from the $r$-process,][]{Bisterzo2011}, as well as in $s$-elements. These stars are called CEMP-$r/s$ stars. The definitions adopted for the various categories of CEMP stars vary for different authors \cite[e.g.][]{BeersChristlieb2005, Jonsell2006, Lugaro2009, Masseron2010, Lee2013}. In our work we adopt the following classification scheme.
\begin{itemize}

\item \emph{CEMP stars} are defined by $[\C/\Fe]>1$. We call ``carbon normal'' those stars with $[\C/\Fe]\leq 1$.

\item \emph{CEMP-$s$ stars} are CEMP stars with $[\Ba/\Fe]>1$ and $[\Ba/\Eu]>0$. The lower limit on [Ba/Eu] is required because high abundances of barium can be produced by the $r$-process, but in that case [Ba/Eu] is negative. \cite{BeersChristlieb2005} propose to use $[\Ba/\Eu]>0.5$, however AGB-nucleosynthesis models, which only include the $s$-process, predict $[\Ba/\Eu]$ close to zero in some circumstances \cite[][]{Lugaro2012, Abate2015-1}, hence our choice of a lower threshold.

\item \emph{CEMP-$r/s$ stars} are CEMP stars with $[\Eu/\Fe]>1$, $[\Ba/\Fe]>1$ and $[\Ba/\Eu]>0$. With this definition, CEMP-$r/s$ stars are a subclass of CEMP-$s$ stars. However, we note that CEMP-$s$ and CEMP-$r/s$ stars possibly have different origins.

\item \emph{CEMP-$r$ stars} are CEMP stars that exhibit $[\Eu/\Fe]>1$ and $[\Ba/\Eu]\leq0$. The upper limit on [Ba/Eu] is to exclude systems that are europium-rich because of the $s$-process, which would show positive [Ba/Eu] \cite[e.g.][]{Abate2015-2, Abate2015-1}.

\item \emph{CEMP-no stars} are CEMP stars with $[\Ba/\Fe]\leq 1$ and $[\Eu/\Fe]\leq 1$.

\end{itemize}

The origin of CEMP-$r/s$ stars is debated. Although currently no study has specifically focused on the orbital properties of CEMP-$r/s$ stars, there are indications that the majority of them belongs to binary systems, as for CEMP-$s$ stars \cite[][]{Lucatello2005a, Hansen2015-4}. However, it is commonly believed that the interiors of AGB stars do not reach sufficiently high neutron densities to activate the $r$-process, and consequently the binary scenario invoked for the formation of CEMP-$s$ stars does not explain the abundances observed in CEMP-$r/s$ stars. Also, CEMP-$r/s$ stars exhibit on average higher abundances of carbon and heavy-$s$ elements (barium, lanthanum and cerium) compared to CEMP-$s$ stars with $[\Eu/\Fe]\leq 1$, whereas the two groups are observed to have approximately the same range of abundances of light-$s$ elements \cite[strontium, yttrium and zirconium,][]{Abate2015-2}.

\cite{Jonsell2006} and \cite{Lugaro2009}, suggest and discuss a variety of formation scenarios. In many of these scenarios the origin of the $s$-elements is pollution from AGB primary stars in binary systems, as for CEMP-$s$ stars, whereas to explain the abundant $r$-elements it has been proposed that: ($i$)~the gas clouds, in which the binary systems were born, were $r$-rich because of supernova explosions of previous-generation stars; ($ii$)~$r$-process nucleosynthesis can be activated under some circumstances in low-metallicity AGB stars; ($iii$)~$r$-elements are ejected by a third, massive star in a hierarchical triple system, or alternatively, in a binary system, by the primary star that, after the AGB evolution, ($iv$)~explodes as Type-1.5 supernovae or ($v$)~undergoes accretion-induced collapse into a neutron star. Alternative formation channels involve radiative levitation of the neutron-capture elements in the stellar atmospheres or self-enrichment during the AGB phase of evolution. %
A quantitative study of the likelihood to form CEMP-$r/s$ stars in these scenarios is as yet missing.

The aim of our paper is to answer the following question: what is the most likely scenario to explain the formation of CEMP-$r/s$ stars? For this purpose we calculate the frequency of CEMP-$r/s$ stars in each of the above formation scenarios, and we compare it with that determined in an observed sample of MP stars.


\section{Observed sample}
\label{data}

Our observational sample is based on the SAGA database of stellar abundances \cite[][last updated August 2015]{Suda2008, Suda2011}, which is arguably the richest collection of heavy-element abundances currently available and includes $1944$ MP stars with $[\Fe/\Hy]<-1$. Stars with higher iron abundances in the SAGA database do not show large enrichments in either barium or europium, and therefore we exclude them from our sample. We select the MP stars with observed abundances of barium and europium, and we ignore systems in which only upper or lower limits are available. In case the chemical composition of a star is reported by multiple sources, we adopt the arithmetic mean of the observed logarithmic abundances. Table \ref{tab:obs} summarises the number of MP stars of different classes included in our sample.

\begin{figure}[!t]
\includegraphics[width=0.48\textwidth]{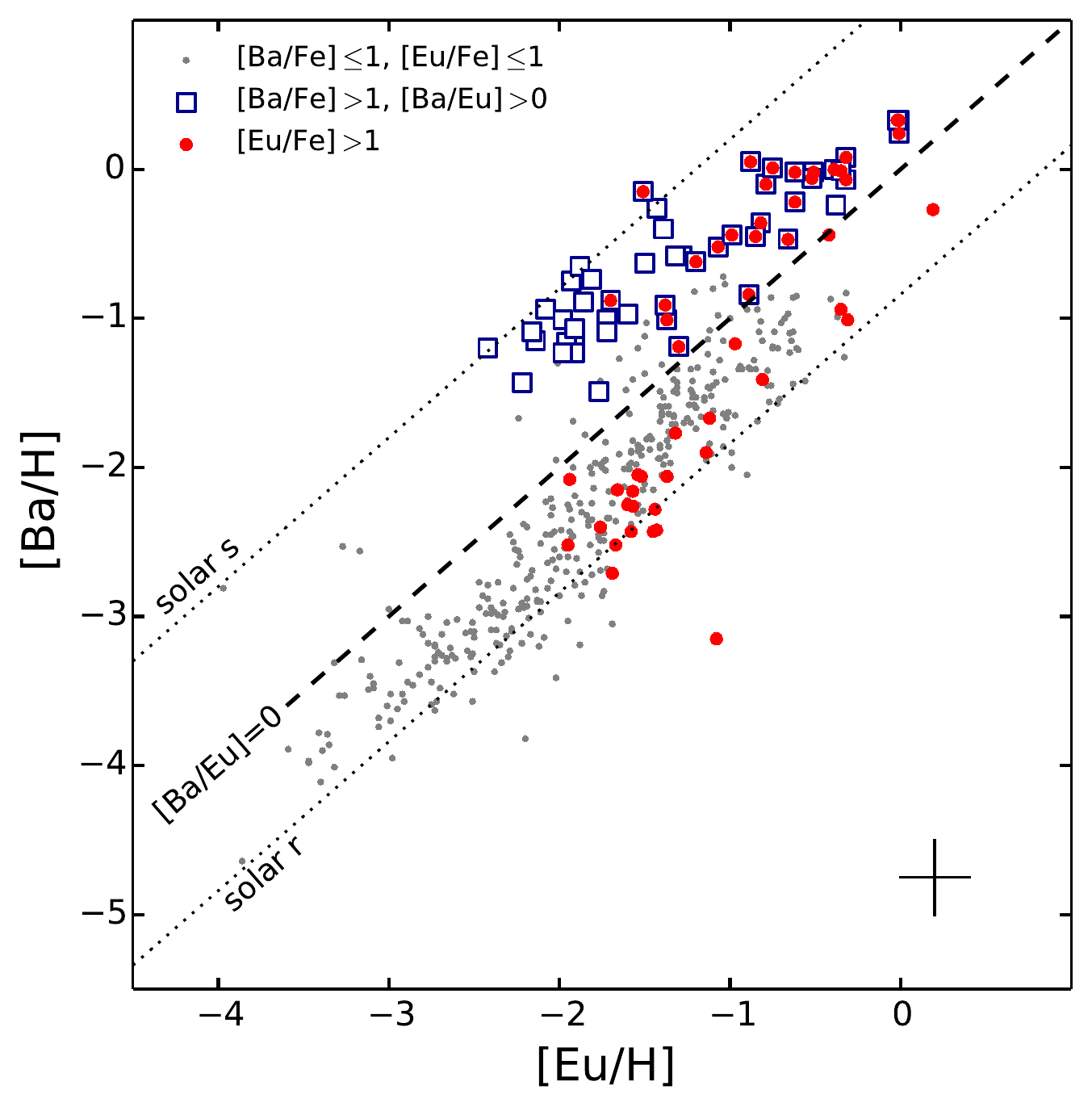}
\caption{[Ba/H] and [Eu/H] of metal-poor stars with $[\Fe/\Hy]<-1$ from the SAGA observational database. Grey and red dots indicate stars with $[\Eu/\Fe]\leq1$ and $[\Eu/\Fe]>1$, respectively. Stars with $[\Ba/\Fe]>1$ and $[\Ba/\Eu]>0$ are shown as blue open squares. The $+$ symbol in the bottom right corner shows the average observational uncertainty. The solar [Ba/Eu] ascribed to the $r$- and $s$-processes are shown as dotted lines \cite[][]{Goriely1999, Masseron2006}. The dashed line represents $[\Ba/\Eu]=0$.}
\label{fig:BaH-vs-EuH}
\end{figure}

Figure \ref{fig:BaH-vs-EuH} shows the $451$ MP stars in our sample with observed abundances of europium and barium, without any assumption about the carbon abundance. Almost all these stars exhibit abundances between the two dotted lines that indicate the barium-to-europium ratios ascribed to the $r$- and the $s$-process in the solar system, $[\Ba/\Eu]_{\odot,r}=-0.84$ and $[\Ba/\Eu]_{\odot,s}=1.2$ respectively \cite[][]{Goriely1999, Masseron2006}. The dashed line separates the majority of stars in our sample, with $[\Ba/\Eu]\leq 0$, from the 77 stars with $[\Ba/\Eu]> 0$. Grey dots indicate the $399$ stars with $[\Ba/\Fe]\leq1$ and $[\Eu/\Fe]\leq1$.  The blue open squares are the $50$ ``$s$-enhanced MP stars'', or MP-$s$ stars, i.e. stars with $[\Ba/\Fe]>1$ and $[\Ba/\Eu]>0$. Stars with $[\Eu/\Fe]>1$ are shown as red circles. Half of these $52$ europium-rich stars have $[\Ba/\Eu]>0$ and are classified as MP-$r/s$ stars in analogy with our definition of CEMP-$r/s$ stars.

The stars in our sample are selected according to their observed abundances of barium and europium, regardless of the carbon abundances. However, $48$ out of $50$ MP-$s$ stars and all MP-$r/s$ stars exhibit $[\C/\Fe]>1$, whereas only five out of $26$ MP-$r$ stars are also carbon-enriched. One of the two MP-$s$ stars that are not classified as CEMP-$s$ has $[\C/\Fe]\approx0.8$ and is a red giant \cite[BD~--01~2582,][]{Simmerer2004, Roederer2014-1}. Hence, in the hypothesis that carbon-rich material was transferred from a companion star, the carbon enhancement could have been reduced by the first dredge-up \cite[e.g.][]{Stancliffe2007, Placco2014}. There currently is no determination of carbon in the literature for the other MP-$s$ star (G~18--24) . Hence, in our sample the proportion of CEMP stars among $s$- and $r/s$-rich MP stars is almost $100\%$, whereas among MP-$r$ stars it is consistent with the overall CEMP/MP fraction (approximately $19\%$).

The frequency of CEMP-$r/s$ stars among CEMP-$s$ stars in our sample is $26/48\approx54\%$. The SAGA database includes $55$ CEMP stars with $[\Ba/\Fe]>1$ and without the europium abundance, that is stars in which an upper limit is determined with $[\Ba/\Eu_{\mathrm{upper}}]>0$ or europium lines are not detected. If we relax our definition of CEMP-$s$ stars to also include these systems, that is we implicitly assume that if europium is not observed its abundance is sufficiently low that $[\Ba/\Eu]>0$, the ratio of CEMP-$r/s$ to CEMP-$s$ stars decreases to approximately $25\%$. The frequency of CEMP-$r/s$ stars provide useful constraints on the theoretical models proposed for the formation of these systems. However, the SAGA database is a collection of stellar abundances published in the literature, and consequently our sample is incomplete and inhomogeneous because of the different properties and selection effects of the original sources.

\begin{table}[!t]
\caption{Classification of stars in our observational sample based on the~SAGA database.}
\label{tab:obs}
\centering
\tiny
\begin{tabular}{ l  r }
\hline
\hline
definition of stellar group	& number\\
							& of stars\\
\hline

$[\Fe/\Hy]<-1$ (MP)																			& 1944	\\
MP with detected barium and europium (including limits)										& 725	\\
$-$ MP with barium and europium abundances													& 451	\\
\hspace{3mm}$-$ MP with $[\Ba/\Fe]>1.0$ 													& 60	\\
\hspace{6mm}$-$ MP-$s$ \hspace{3mm}($[\Ba/\Fe]>1.0$ and $[\Ba/\Eu]>0.0$)					& 50	\\
\hspace{6mm}$-$ MP-$r/s$ ~($[\Ba/\Fe]>1.0$, $[\Ba/\Eu]>0.0$, and $[\Eu/\Fe]>1.0$)			& 26	\\
\hspace{6mm}$-$ MP-$r$ \hspace{3mm}($[\Eu/\Fe]>1.0$, $[\Ba/\Eu]\leq0.0$)					& 26	\\
\hline
MP with carbon abundances																	& 947	\\
$-$ MP with $[\C/\Fe]\leq 1.0$																& 763	\\
$-$ MP with $[\C/\Fe]>1.0$ (CEMP)															& 184	\\
\hspace{3mm}$-$ CEMP-no$^\dag$ 																& 75	\\
\hspace{3mm}$-$ CEMP with barium abundances													& 152	\\
\hspace{6mm}$-$ CEMP with $[\Ba/\Fe]>1.0$ without europium abundances						& 55	\\
\hspace{6mm}$-$ CEMP with barium and europium abundances									& 56	\\
\hspace{9mm}$-$ CEMP-$r$																	& 5		\\
\hspace{9mm}$-$ CEMP with $[\Ba/\Fe]>1.0$ (with europium abundances)						& 51	\\
\hspace{12mm}$-$ CEMP-$s$				& 48	\\
\hspace{15mm}$-$ CEMP-$r/s$			& 26	\\

\hline
\end{tabular}
\tablefoot{$^\dag$CEMP stars with $[\Ba/\Fe]\leq 1$ and $[\Eu/\Fe]\leq 1$, or without detections of barium and europium.\\}
\end{table}

\section{Results}
\label{results}

Several scenarios have been proposed in the literature to explain the formation of CEMP-$r/s$ stars \cite[e.g.][]{Hill2000, Cohen2003, Zijlstra2004, Ivans2005, Jonsell2006, Lugaro2009, Herwig2011, Dardelet2015}. To determine the likelihood of these hypotheses, we calculate the frequency of CEMP-$r/s$ stars predicted in each scenario and we compare it with the observed proportion of CEMP-$r/s$ stars among CEMP-$s$ stars.

\subsection{Pre-enrichment in $r$-rich material and pollution from an AGB star in a binary system}
\label{pre-enrich}

In this scenario, the enrichment in $r$- and $s$-elements in CEMP-$r/s$ stars originate from two independent sources. The CEMP-$r/s$ star would be the secondary star in a binary system born in a gas cloud which was pre-enriched in $r$-elements, for example because of the explosion of a nearby supernova. The observed carbon and $s$-elements are instead produced by the primary star during the AGB phase, and subsequently transferred by stellar winds onto the low-mass companion, which is the star observed today (e.g. \citealp{Jonsell2006}, \citealp{Bisterzo2011}, and references therein).

To investigate this scenario, we use the population synthesis model developed by \cite{Izzard2004, Izzard2006, Izzard2009, Izzard2010} to simulate a population of binary stars in which the abundances of $r$-process elements are initially enhanced by a random factor. Our model population is constructed according to the default model set A of \cite{Abate2015-3}. Our simulated grid consists of approximately $1.8$ million binary systems with primary and secondary masses in the intervals $[0.5,\,8]\,\Msun$ and $[0.1,\,0.9]\,\Msun$, respectively, orbital separations $a_i$ between $50$ and $5\times10^6\,\Rsun$, and seven different values of the mass of the partial-mixing zone, a parameter that determines the abundance distribution of neutron-capture elements synthetised in the AGB phase \cite[][]{Karakas2010, Lugaro2012, Abate2015-1}. The distribution of primary masses in our population follows the solar-neighbourhood initial mass function (IMF) proposed by \cite{Kroupa1993}, whereas all secondary masses are equally likely and the separation distribution is flat in $\log a_i$. We compute the wind mass-transfer process according to a wind Roche-lobe overflow model described by \citet[Eq. 9]{Abate2013}, and adopting the spherically-symmetric-wind approximation. We assume that the transferred material is mixed throughout the accreting star to mimic the effect of thermohaline mixing, which is expected to be efficient in low-mass stars \cite[][]{Stancliffe2007, Stancliffe2008}.

As initial composition we adopt the solar abundance distribution of \cite{Asplund2009} scaled down to metallicity $10^{-4}$ (that is $[\Fe/\Hy]\approx-2.2$). Our method to simulate a random pre-enrichment in barium and europium produced by the $r$-process is summarised as follows. For each stable isotope $i$ of barium and europium, we take from \cite{Bisterzo2011} the fraction of its solar abundance which is produced by the $r$-process, $f_{i,r}$. We multiply our initial, solar-scaled abundances, $X_i$, of each isotope times their corresponding fraction $f_{i,r}$. We sum the abundances of the isotopes of the same element to calculate the total abundances of barium and europium ascribed to the $r$-process, and we calculate the decimal logarithm of these two values. Subsequently, for each synthetic binary system we select one value from a randomly generated sequence of $1.8$ million numbers uniformly distributed between $-1$ and $3$, and we add this number to the logarithmic abundances of barium and europium to simulate an initial enhancement (or depletion) of the $r$-component of these two elements. The interval in which the random numbers are generated reproduces the range of variation of $[\Eu/\Fe]$ in our observed~sample. As a consequence of this artificial pre-pollution, the initial model distributions of $[\Eu/\Fe]$ and $[\Ba/\Fe]$ are flat over the abundance ranges $[-1,\,3]$ and $[-0.2,\,2.2]$, respectively.

We evolve our binary systems with these initial conditions and we determine the number of stars that are visible as a function of their luminosities according to the selection criteria of \citet[Sect. 2.3]{Abate2015-3} with $V$-magnitude limits at $6$ and $16.5$. In Fig. \ref{fig:EuFe-BaFe} we show the distributions of barium and europium relative to iron for our synthetic population. Carbon-rich stars with $[\C/\Fe]>1$ are shown in red, with darker colours representing regions of higher probability, while carbon-normal stars are shown in blue. Figure~\ref{fig:EuFe-BaFe} also shows the abundances of observed CEMP and carbon-normal metal-poor stars (red crosses and blue dots, respectively). Because our model is computed for $[\Fe/\Hy]=-2.2$, we restrict our observed sample to stars with iron abundance within a factor of five from the model value, $-2.9\leq [\Fe/\Hy] \leq-1.5$. This selection leaves us with $49$~CEMP, $43$~CEMP-$s$, and $22$~CEMP-$r/s$ stars, that is more than $80\%$ of our entire carbon-rich sample. 

The theoretical and observed abundances differ in many aspects. 
\begin{enumerate}
\item Observed carbon-rich and carbon-normal stars appear as two distinct and clearly separated populations. We compute the linear regression of the observed europium and barium abundances weighted with the observational errors, and we determine the following equations for the ratio of barium to europium in carbon-normal and CEMP-$s$ stars, respectively, 
\begin{align}
	[\Ba/\Fe] &= (0.8 \pm 0.1)\times [\Eu/\Fe] - (0.4 \pm 0.1)~~,	\label{eq:fitVMP}\\
	[\Ba/\Fe] &= (0.7 \pm 0.1)\times [\Eu/\Fe] + (1.1 \pm 0.1)~~,	\label{eq:fitCEMP-s}
\end{align}
represented in Fig. \ref{fig:EuFe-BaFe} with blue-dashed and magenta-solid lines. If we compute the linear fit for CEMP-$s$ stars with $[\Eu/\Fe]\le1$ and CEMP-$r/s$ stars separately we find almost identical results as in Eq.~\ref{eq:fitCEMP-s}. Equations \ref{eq:fitVMP} and \ref{eq:fitCEMP-s} show that the empirical relation between [Eu/Fe] and [Ba/Fe] has almost the same slope in the two classes of stars, and CEMP-$s$ stars have [Ba/Fe] on average $1.5$ dex higher than carbon-normal stars at the same [Eu/Fe]. The relations described by Eqs. \ref{eq:fitVMP} and \ref{eq:fitCEMP-s} are likely consequences of the barium-to-europium ratios produced in the $r$- and in the $s$-process, respectively.
In contrast, in our simulation there is not a clear separation between the two groups (colour distributions in Fig.~\ref{fig:EuFe-BaFe}). Because $r$- and $s$-enrichments are independent, the initial [Eu/Fe] and [Ba/Fe] follow approximately the relation described by Eq. \ref{eq:fitVMP}, whereas the final abundances depend on the amount of material transferred from the AGB primary stars onto their secondary companions. Higher accreted masses cause stronger abundance enhancements of carbon and barium. Because AGB stars generally do not produce high europium abundances, [Eu/Fe] is essentially determined by its initial value, except for stars with initial [Eu/Fe] less than about $0.5$.
Consequently, for each [Eu/Fe], which is given by the initial $r$-enrichment, we find stars with all barium abundances up to two dex higher than predicted by a pure $r$-process, and the correlation between [Eu/Fe] and [Ba/Fe] described by Eq. \ref{eq:fitCEMP-s} is completely washed out. \cite{Lugaro2009, Lugaro2012} and \cite{Abate2015-3} criticise this scenario of independent $r$- and $s$-process contributions with analogous arguments.
%

\begin{figure}[!t]
\centering
\includegraphics[width=0.5\textwidth]{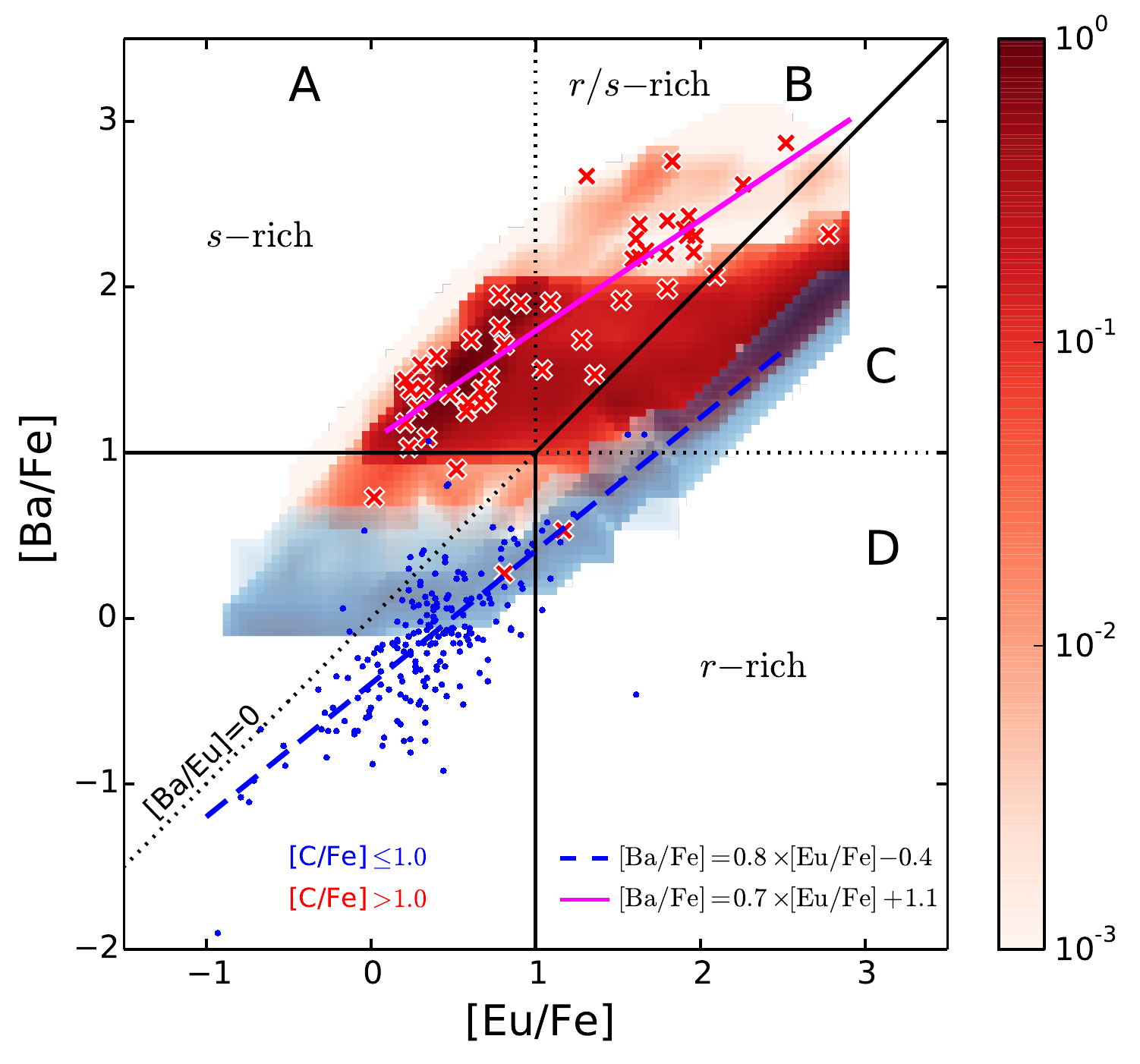}
\caption{[Ba/Fe] vs [Eu/Fe] in carbon-normal and carbon-rich metal-poor stars (blue dots and red crosses, respectively). Stars enriched in $s$-elements occupy the region above the horizontal and diagonal solid black lines, corresponding to zones A and B of the plot. According to our adopted definitions, CEMP-$r/s$ stars are in zone B. In zone C there are stars with $[\Ba/\Fe]>1$, $[\Eu/\Fe]>1$, and $[\Ba/\Eu]<0$. CEMP-$r$ stars and carbon-normal stars with $[\Eu/\Fe]>1$ and $[\Ba/\Fe]\leq1$ are in zone D. The dashed-blue and solid-magenta lines are linear fits of the carbon-normal stars and CEMP-$s$ stars, respectively. The red and blue distributions represent our the results of our theoretical model for CEMP and carbon-normal metal-poor stars.}
\label{fig:EuFe-BaFe}
\end{figure}

\item Only two of the $205$ carbon-normal stars in our metal-poor sample have abundances of europium and barium relative to iron higher than one (CS~$29497$--$004$ and CS~$31082$--$001$, in section C of Fig.~\ref{fig:EuFe-BaFe}, have $[\Eu/\Fe] = 1.7$ and $1.6$, respectively, and $[\Ba/\Fe]=1.1$). In contrast, $12$ CEMP-$r/s$ stars have $[\Eu/\Fe]>1.7$ up to $[\Eu/\Fe] = 2.5$. If the $r$- and $s$-enrichments are independent, we expect to observe carbon-normal MP stars and CEMP-$r/s$ stars with europium abundances approximately in the same range. 
\item The observed fraction of carbon-normal stars with $[\Eu/\Fe]>1$, $[\Ba/\Fe]>1$, and $[\Ba/\Eu]\leq 1$ (in zone C of Fig. \ref{fig:EuFe-BaFe}) is $2/205\approx1\%$. In contrast, approximately $28\%$ of our simulated carbon-normal stars are in zone C of the plot. This discrepancy is unlikely to be caused by an observational bias against carbon-normal, $r$- and $s$-rich metal-poor stars. The spectra of metal-poor stars are selected evaluating the strength of the calcium and iron lines, and these metallicity indicators are not perturbed by the presence of barium or europium.
\item Similarly, the fraction of synthetic CEMP stars in zone C of the plot is $42\%$. In contrast, only two of the $49$ observed CEMP stars are in this region, that is approximately $4\%$. The high fraction of synthetic carbon-rich and carbon normal stars in zone C of Fig. \ref{fig:EuFe-BaFe} is a consequence of our assumptions on the initial abundances of barium and europium. To reduce this fraction, it is necessary to constrain the initial barium abundances within $[\Ba/\Fe]<1$, that corresponds to europium abundances in $[\Eu/\Fe]\lesssim1.8$ \cite[][]{Goriely1999, Pagel2009}. Consequently, if we adopt these initial assumptions in our model, the europium abundances of eight observed CEMP-$r/s$ stars with $[\Eu/\Fe]>1.8$ are not reproduced. Alternatively it would be necessary to assume that the barium-to-europium ratio of a pure $r$-process at low metallicity is $[\Ba/\Eu]\approx-2$, but this hypothesis is not borne out by the observations (cf. Fig. \ref{fig:BaH-vs-EuH}).
\item In our simulation the proportion of CEMP-$r/s$ among CEMP-$s$ stars is approximately $22\%$, and it underestimates the observed fraction of about $51\%$ by more than a factor of two. If we also include CEMP stars with $[\Ba/\Fe]>1$ and an upper limit on europium such that $[\Ba/\Eu]>0$, and we assume that all barium-enhanced CEMP stars with no europium detection are CEMP-$s$ stars, the number of observed CEMP-$s$ stars increases to $86$ and the observed ratio of CEMP-$r/s$ to CEMP-$s$ stars is $26\%$, that is approximately consistent with the simulations. However, even in this case, there is a large discrepancy in the barium-abundance distributions of observed and modelled CEMP-$r/s$ stars. The majority of the observed CEMP-$r/s$ stars ($16/22\approx73\%$) are strongly barium-enhanced and show $[\Ba/\Fe]>2$. In contrast, the cumulative fraction of synthetic CEMP-$r/s$ stars with barium abundance $[\Ba/\Fe]> 2$ is $4\%$, almost a factor of twenty less than the observations. Hence, this model does not adequately reproduce the observed distribution of barium abundances in CEMP-$r/s$ stars.

\end{enumerate}

These discrepancies also arise if we modify our model assumptions. The abundance distributions of barium and europium are essentially independent from the assumptions about the efficiencies of the wind-accretion process and the angular-momentum loss, and the initial distributions of secondary masses and orbital separations \cite[][]{Abate2015-3}. An IMF weighted towards intermediate masses favours stars that typically produce higher barium abundances than low-mass stars with $M<1.5\Msun$ \cite[][]{Abate2015-3}, but the proportion of CEMP-$r/s$ stars with $[\Ba/\Fe]>2$ is low (less than $\approx5\%$). If we assume that the accreted material remains at the surface of the secondary stars until is mixed by the first dredge-up, as in model set D of \cite{Abate2015-3}, we find the maximum frequency of CEMP-$r/s$ stars with $[\Ba/\Fe]>2$, that is about $84\%$, higher than the observations. However, this model does not reproduce the abundance distributions of carbon and $s$-elements in CEMP stars \cite[cf. model D of][]{Abate2015-3} nor the observed correlation between barium and europium abundances, analogously to our default model. The discrepancy between the small number of stars observed in zone C of Fig. \ref{fig:EuFe-BaFe} and the much higher frequencies of synthetic carbon-normal and carbon-rich MP stars is a consequence of the adopted initial abundances of barium and europium, and therefore does not depend on the model set.

\subsection{Binary mass-transfer scenario with \emph{i}-process nucleosynthesis in AGB stars}
\label{i-proc}

In this scenario, the AGB star invoked to explain the enrichment in carbon and $s$-elements is also responsible for the production of $r$-elements. \cite{Cohen2002} and \cite{Jonsell2006} dismiss this channel with the argument that in AGB stars it is not possible to create the high neutron flux required for $r$-process nucleosynthesis. However, recent simulations show that relatively high neutron densities, between $10^{12}$ and $10^{15}\,\cm^{-3}$, can be reached in very low-metallicity AGB stars \cite[][]{Campbell2008, Cristallo2009-2, Campbell2010, Stancliffe2011}, in the very-late thermal pulse of post-AGB stars \cite[][]{Herwig2011, Bertolli2013, Herwig2014, Woodward2015} or during the thermal-pulse evolution of low-metallicity super-AGB stars \cite[][]{Jones2015}. At these densities, which are in-between $10^8\,\cm^{-3}$ and $10^{23}\,\cm^{-3}$ (values normally adopted in $s$- and $r$-process models, respectively, \citealp{Pagel2009}), the \emph{intermediate} neutron-capture process ($i$-process, \citealp{Cowan1977}) is triggered. 

\cite{Hampel2015} use a one-zone model to simulate the properties of a gas with temperature, density and chemical composition as in the intershell region of an AGB star and calculate the nucleosynthetic products in the presence of neutron densities between $10^{12}$ and $10^{16}\,\cm^{-3}$. The author shows that the $i$-process predicts high abundances of elements traditionally associated with the main $s$-process component (e.g. barium, lanthanum, cerium, and lead) and with the $r$-process (e.g. iodine and europium), while it does not significantly modify the abundances of elements normally associated to the weak $s$-process (strontium, yttrium and zirconium). 
These abundance patterns are typical of CEMP-$r/s$ stars, as discussed for example by \cite{Abate2015-2} and \cite{Hollek2015}. With her model, \cite{Hampel2015} is able to reproduce the surface abundances of $20$ observed CEMP-$r/s$ stars, significantly improving the results obtained by \cite{Abate2015-2} with a standard AGB-nucleosynthesis model.

To investigate the likelihood of this scenario it is necessary to know under what circumstances AGB stars experience thermal pulses characterised by high neutron fluxes. This information is very uncertain. Proton-ingestion episodes with large neutron production are found in models of very low-metallicity ($Z\leq 10^{-4}$), low-mass ($M\leq 2\Msun$) stars \cite[e.g.][]{Campbell2008, Cristallo2009-2, Stancliffe2011}. At present, it has not been investigated whether these episodes occur in the whole metallicity range of CEMP-$r/s$ stars, i.e. up to $Z\approx10^{-3}$. In addition, proton ingestion is triggered in the early thermal pulses of the AGB stars. It is currently unclear whether the subsequent nucleosynthesis dominates the surface abundances at the end of AGB evolution, when most of the envelope mass is ejected, or whether traces of the $i$-process remain detectable \cite[][]{Campbell2008, Cristallo2009-2}.

Super-AGB stars are normally considered to form at masses higher than $6\,\Msun$. Because AGB stars of these masses experience hot-bottom burning \cite[e.g.][]{Boothroyd1993, Izzard2007}, the material accreted from these stars is expected to be in most cases nitrogen-rich, rather than carbon-rich \cite[][]{Pols2012}. Also, such massive stars are disfavoured by the IMF, and therefore the theoretical ratio of CEMP-$r/s$ to CEMP-$s$ stars is lower than the observations: about $0.04\%$ with a solar-neighbourhood IMF. This fraction increases if we adopt an IMF weighted towards intermediate-mass stars, for example that proposed by \cite{Lucatello2005a}, which is a gaussian function with $\mu_{\log_{10} M} = 0.79$ and $\sigma_{\log_{10} M} = 1.18$. However, the probability of making a CEMP star in a binary system with a $M\ge6\Msun$ donor star is about $2\%$, more than an order of magnitude lower than the observations. %
A similar argument constrains the conditions to trigger the very-late thermal pulse in post-AGB stars. If these conditions are mass-dependent, a relatively low threshold mass is necessary. To produce a proportion of CEMP-$r/s$ stars of about $30\%$ in this scenario with a solar-neighbourhood IMF, the high neutron exposures should be reached in all stars of initial mass down to $1.2\,\Msun$. Further detailed calculations are necessary to investigate if the $i$-process can be activated at such low masses.

\subsection{Triple systems with SN explosion and AGB pollution}
\label{triple}

In this scenario CEMP-$r/s$ stars are formed in triple systems in which the tertiary, least massive star accretes $r$-elements from the ejecta of the supernova explosion of the primary star, and carbon and $s$-elements from the wind of the secondary star in the AGB phase of evolution. 
To maximise the ratio of CEMP-$r/s$ to CEMP-$s$ stars produced in this scenario, we make the following, rather unrealistic, assumptions: ($i$) all stars form in hierarchical triple systems, ($ii$) intermediate-mass, low-metallicity AGB stars have negligible mass loss, as argued by \cite{Wood2011}, and consequently all stars of mass $M \geq 3\Msun$ explode as supernovae, ($iii$) the accretion of material onto the low-mass component of the triple system is sufficiently efficient to enhance europium above the threshold $[\Eu/\Fe]=1$, regardless of the orbital separation of the primary star. 

Under these assumptions, the necessary condition to form a CEMP-$r/s$ star is that the primary star in the triple system is more massive than $3\,\Msun$. If we adopt the solar-neighbourhood IMF proposed by \cite{Kroupa1993}, the probability of forming such a primary star is $P(M\geq 3\Msun)=0.014$, that is, more than a factor of ten lower than the observations. If we adopt the IMF weighted towards intermediate-mass stars proposed by \cite{Lucatello2005a}, the probability of a star more massive than $3\Msun$ is $P(M\geq 3\Msun)=0.31$, that is within the range of uncertainty of the observations. However, we note that \cite{Lucatello2005a} proposed their IMF for stars of extremely low metallicity, $[\Fe/\Hy]<-2.5$, and similarly the suppression of the mass loss proposed by \cite{Wood2011} was only found in models at $[\Fe/\Hy]=-4.2$, whereas the most iron-deficient CEMP-$r/s$ star in our sample has $[\Fe/\Hy]=-3.1$ and $15$ of our $26$ CEMP-$r/s$ stars have $[\Fe/\Hy]>-2.5$.

These results are obtained assuming that all stars are formed in triple systems. The multiples-to-binaries ratio estimated by \cite{Rastegaev2010} for MP stars is $10/64\approx0.16$. Taking into account this factor, with the IMF of \cite{Lucatello2005a} the CEMP-$r/s$ frequency is reduced to approximately $5\%$, that is a factor of five up to ten times lower than in our observed sample, depending on whether we count all CEMP stars without the europium abundance and $[\Ba/\Fe]>1$ (cf. Tab. \ref{tab:obs}). This proportion decreases even further if we consider that mass accretion from supernova ejecta is typically inefficient, unless the orbital separation is short \cite[][]{Liu2015}. In contrast, the initial period of the inner binary system has to be at least $1,\!000$ days, because the tertiary star has to transfer carbon and $s$-elements from the secondary AGB star, and binary systems in closer orbits enter in a common envelope without mass accretion. In stable triple systems the orbit of the primary star is typically much wider than that of the inner binary system \cite[][]{Kiseleva1994}. The calculations of \cite{Liu2015} predict that at orbital periods longer than $1,\!000$ days the transferred mass is less than $10^{-5}\,\Msun$. Consequently, if the transferred material is diluted in a layer as thin as $0.01\,\Msun$ in the envelope of the accreting star, the amount of transferred material is not sufficient to enhance the europium abundances above $[\Eu/\Fe]=1$, even if the supernova ejecta exhibit $[\Eu/\Fe]=3$. Hence, this formation scenario of CEMP-$r/s$ stars is considered implausible.

\subsection{Binary star with $1.5$-supernova pollution}
\label{1.5SN}

In this formation scenario the CEMP-$r/s$ star would be the secondary star of a binary system in which the primary star produced the $s$-elements during the AGB phase and subsequently exploded as a Type 1.5 supernova producing the $r$-elements. It is therefore necessary to assume that the mass loss along the AGB is sufficiently high to allow significant accretion of $s$-elements, but not too high otherwise the core of the star does not reach the Chandrasekhar mass. \cite{Zijlstra2004} propose that low-metallicity AGB stars with initial mass higher than about $3-4\,\Msun$ undergo this evolution. These AGB stars should also produce carbon and $s$-elements. According to the simulations of \cite{Abate2015-3} the proportion of carbon-enhanced stars that are formed in binary systems with primary masses higher than $3\,\Msun$ is less than $1\%$ if a solar-neighbourhood IMF is adopted, and increases to about $14\%$ with the IMF of \cite{Lucatello2005a}. Simulations of the interactions of core-collapse supernova ejecta with a main-sequence companion star show that the amount of accreted mass critically depends on the binary separation, and it is less than $10^{-4}\,\Msun$ for a $0.9\,\Msun$ at periods longer than about $40$ days \cite[][]{Liu2015}. The cumulative fraction of CEMP stars formed in closer orbits with the IMF of \cite{Lucatello2005a} is less than $5\%$, and consequently the ratio of CEMP-$r/s$ to CEMP-$s$ stars is less than~$0.7\%$. In addition, according to many authors the explosion of a Type 1.5 supernova destroys the donor star \cite[][]{Nomoto1976, Iben1983, Lau2008}, although \cite{Arnett1974} argues that a remnant may survive in some cases, and therefore the binary system is disrupted. In contrast, many CEMP-$r/s$ stars are observed in binary systems \cite[][]{Lucatello2005a, Hansen2015-4} and therefore could not have originated by this formation channel.

\subsection{Binary system with AGB- and AIC-pollution}
\label{AIC}

This scenario involves two phases of mass transfer. In the first, a relatively massive primary AGB star \cite[$M \approx 3-12\,\Msun$ according to][]{Cohen2003} transfers carbon and $s$-elements onto its low-mass ($M_*<0.9\,\Msun$) companion, and later ends its life as a white dwarf. Subsequently, the secondary star transfers some material onto the white dwarf which collapses into a neutron star. This scenario is supported by the study of \cite{Qian2003} on r-element production in neutrino winds associated to accretion-induced collapse (AIC).

We consider this scenario unlikely for the following reasons. First, CEMP stars formed in binary systems with primary masses above $3\,\Msun$ are unlikely, as also discussed in sections \ref{triple} and \ref{1.5SN}. We also note that AIC models have mostly investigated the collapse of oxygen-neon white dwarfs, which descend from stars more massive than $5\,\Msun$ \cite[e.g.][]{Qian2003, Dessart2007}, for which the probability of forming a CEMP star is less than $5\%$ even with the IMF of \cite{Lucatello2005a}. 

Secondly, the $s$-element distribution predicted for relatively massive AGB stars ($M\ge3\,\Msun$) is not consistent with the abundances observed in CEMP-$s$ stars, which are mostly reproduced by stellar models of mass $M<2\,\Msun$ \cite[e.g.][]{Bisterzo2011, Bisterzo2012, Abate2015-1, Abate2015-2}. Also, the $r$-process nucleosynthesis in AIC events is largely uncertain \cite[][]{Qian1996, Qian2003}, therefore it is unclear whether in this scenario the total amount of $r$-elements produced and their abundance distribution reproduce the chemical composition of observed CEMP-$r/s$ stars.

Third, this scenario works in a narrow range of orbital separations, because at the end of the first mass transfer the secondary star needs to be close enough to the white dwarf to subsequently fill its Roche-lobe and undergo the second mass-transfer phase. A star of $0.9\,\Msun$ and a white dwarf of $1\,\Msun$ need to be in an orbit of less than about $200$ days to undergo Roche-lobe overflow during the main-sequence or the red-giant phase. This star should be already carbon-enhanced because the AIC event is responsible of the $r$-pollution. According to the simulations of \cite{Abate2015-3}, the cumulative probability of CEMP stars in binary systems with orbital periods less than $200$ days is about $8\%$ at maximum. Hence, the frequency of CEMP-$r/s$ stars is reduced by this factor, and consequently the ratio of CEMP-$r/s$ to CEMP-$s$ stars is much lower than observed.

\subsection{Radiative levitation}
\label{radlev}

In stars, elements with high atomic mass are normally believed to sink more than light elements because of gravitational settling. However, partially-ionized heavy elements usually have large photon-absorption cross sections, and hence they can be accelerated outwards by radiative pressure. This process, known as radiative levitation, may cause the abundance of heavier elements to increase towards the surface.
Consequently, the~overabundances of neutron-capture elements detected in CEMP-$r/s$ stars could in principle be explained by radiative levitation, although this process has never been studied for elements heavier than nickel at low metallicities \cite[][]{Richard2002-3}.

In stars, the efficiency of radiative levitation increases with decreasing convective-envelope mass and increasing effective temperature. Consequently, this process is most efficient in main-sequence stars close to the turnoff and Hertzsprung-gap stars, whereas its effect is negligible in giants (\citealp{Richard2002-1}, Matrozis et al. in prep.). However, the observed CEMP-$r/s$ stars are not preferentially main-sequence or turnoff stars. Only $15$ out of $26$ CEMP-$r/s$ stars in our sample have surface gravities in the range $3.5<\loggunits<4.25$ and effective temperatures higher than $\Teff=5,\!800\,$K, and consequently exhibit surface abundances that have been possibly modified by radiative levitation. In contrast, $11$ CEMP-$r/s$ stars ($42\%$ of our sample) have $\loggunits<3.5$, hence radiative levitation cannot be generally invoked to explain the enrichments in neutron-capture elements observed in CEMP-$r/s$ stars. In addition, this scenario does not explain the differences in abundances between CEMP-$s$ and CEMP-$r/s$ stars.

\subsection{Pre-enrichment in $r$-rich material and self pollution}

In this scenario, the CEMP-$r/s$ star formed in an environment which was enriched in $r$-elements, for example because of an early supernova, and it self-enriched its surface with carbon and $s$-elements during the AGB phase of evolution \cite[][]{Hill2000, Cohen2003}. However, as mentioned in Sect. \ref{radlev}, more than half of the CEMP-$r/s$ stars in our sample have $\logg>3.5$ and have not reached the giant phase yet. This hypothesis is also dismissed by \cite{Jonsell2006} with similar arguments. In addition, if the $r$- and $s$-enrichments are independent, the observed correlation between europium and barium abundances is not reproduced by the models, as discussed in Sect. \ref{pre-enrich}. Also, this scenario does not explain the high frequency of binary systems detected among CEMP-$r/s$ stars \cite[][]{Lucatello2005a, Hansen2015-4}.


\section{Discussion}
\label{discussion}

The results presented in Sect. \ref{results} show that all the scenarios proposed so far to explain the formation of CEMP-$r/s$ stars have difficulties in reproducing the properties of the observed CEMP-$r/s$ population, in particular the proportion of these systems among CEMP-$s$ stars. Scenarios that involve multiple phases of mass transfer, either in binary or in triple systems, underestimate this proportion because the $r$-elements are produced by the explosion or collapse of a relatively massive star ($M>3\,\Msun$), which are rarely formed in a solar-neighbourhood IMF and also, in most cases, even assuming an IMF weighted towards intermediate-mass stars. In addition, impact simulations of supernova explosions on binary companions show that the mass accreted from supernova ejecta is typically lower than $10^{-4}\,\Msun$ for orbital periods longer than about $100$ days \cite[e.g.][]{Pan2012, Liu2015}. Because the fraction of CEMP stars that are formed in such close orbits is low \cite[][]{Abate2015-3}, very few CEMP-$r/s$ stars are consequently formed via these channels. Also, it is currently unclear whether the $r$-element distribution produced in supernovae would reproduce the abundances of observed CEMP-$r/s$ stars \cite[][]{Arcones2011}.

A frequency of CEMP-$r/s$ stars that approaches the observations is predicted in the hypothesis of independent enrichments in $s$- and $r$-elements, in which the latter are the result of pre-pollution of the gas in which CEMP-$r/s$ stars were born. However, models based on this hypothesis fail to reproduce the observed correlation between the barium and europium abundances in CEMP-$s$ stars (as also discussed e.g. by \citealp{Lugaro2009, Lugaro2012}, and \citealp{Abate2015-3}) and they predict too many carbon-normal stars with high abundances of europium and barium. \cite{Lugaro2012} also note that in stellar-nucleosynthesis models the final europium abundance is essentially independent of its value at the beginning of the AGB evolution, unless the initial enrichment is higher than $[r/\Fe]\approx1.5$. Hence, lower initial enhancements are essentially washed out by the $s$-process nucleosynthesis occurring along the AGB. Also, the fact that the average abundance of heavy-$s$ elements (barium, lanthanum, cerium) is more than about $0.9$ dex higher than that of the light-$s$ elements (strontium, yttrium, zirconium), i.e. one of the characteristics of most CEMP-$r/s$ stars which is not predicted by AGB models \cite[][]{Abate2015-2, Hollek2015}, can only be reproduced if high initial $r$-enhancements are assumed ($[r/\Fe]\geq 1.5$, e.g. \citealp{Bisterzo2012}). 

Recent simulations suggest that neutron densities sufficiently high to trigger the $i$-process ($10^{12}-10^{16}\,\cm^{-3}$ approximately) are reached if hydrogen-rich material is injected in region processed by helium burning. These proton-ingestion episodes are found to occur during the early pulses of very low-metallicity, low-mass AGB stars \cite[][]{Campbell2008, Cristallo2009-2, Stancliffe2011}, in the late thermal pulses of super-AGB stars \cite[][]{Jones2015}, and in the very-late thermal pulse of post-AGB stars \cite[][]{Herwig2011, Herwig2014}. The abundance patterns predicted by nucleosynthesis models at these neutron densities reproduce the surface abundances observed in many CEMP-$r/s$ stars (\citealp{Dardelet2015}, \citealp{Hampel2015}, Hampel et al. in prep.). If the preliminary results of these theoretical models are confirmed, AGB stars that fulfill the conditions to undergo the $i$-process would be the most promising candidates to explain the CEMP-$r/s$ abundances in the context of the binary mass-transfer scenario which is also invoked for the formation of CEMP-$s$ stars. This scenario is also consistent with the observational evidence that many CEMP-$r/s$ stars are found in binary systems \cite[][]{Abate2015-1, Hansen2015-4}, although a detailed analysis (similar to the studies on CEMP-$s$ stars performed by \citealp{Lucatello2005a}, \citealp{Starkenburg2014}, and \citealp{Hansen2015-4}) focused on the orbital properties of these objects is currently missing. 

The $i$-process is also invoked to explain the abundances observed in post-AGB stars in the Milky Way (such as the Sakurai's Object, \citealp{Herwig2011}) and in the Magellanic Clouds \cite[][]{Lugaro2015}, which in some cases have very different abundance distributions than CEMP-$r/s$ stars. For example, some post-AGB stars have been observed to be lead deficient \cite[e.g.][]{DeSmedt2014} and with light-$s$ elements much more enhanced than heavy-$s$ elements (e.g. \citealp{Herwig2011}, and \citealp{Jones2015}). In contrast, all CEMP-$r/s$ stars with observed lead abundances in our sample have $[\Pb/\Fe]\geq 2.5$, and positive ratios of heavy-$s$ to ligh-$s$ elements, $[\hs/\ls]\geq 0.5$. Further work is necessary to calculate if the range of element ratios produced in $i$-process nucleosynthesis spans over the wide range observed in CEMP-$r/s$ stars and post-AGB stars.

To determine whether within this ``$i$-process scenario'' the properties of CEMP-$r/s$ stars are reproduced, the following aspects need to be clarified.
\begin{itemize}
\item Following on the work of \cite{Herwig2011, Herwig2014}, \cite{Stancliffe2011}, and \cite{Woodward2015}, more three-dimensional hydrodynamical simulations are necessary to study the conditions under which proton-ingestion events occur in the helium flash of AGB stars. These simulations should precisely determine the intervals of neutron densities and neutron exposures that are produced as a function of the temperature, density, and metallicity of the stellar layers in which the proton ingestion takes place.
\item Based on the results of these simulations, detailed stellar-evolution models should be used to investigate how likely it is for AGB stars of different masses and metallicities to experience a proton-ingestion event, at what stage of the evolution and how many times this event can occur, and what are its consequences on the subsequent evolution of the star.  Also, it is necessary to study whether the nucleosynthetic products of the proton-ingestion event are actually mixed to the surface, and whether this occurs when the star is undergoing substantial mass loss.
\item Nucleosynthesis models are required to determine whether at the range of neutron exposures predicted by the detailed hydrodynamical simulations the production of neutron-capture elements reproduce the abundances observed in CEMP-$r/s$ stars. The preliminary results of \cite{Dardelet2015}, \citet[][]{Hampel2015}, and Hampel et al. (in prep.) support this hypothesis. The dependence of the abundance distribution on the temperature, density and chemical composition of the gas undergoing the neutron-capture process needs to be investigated to determine the ranges of abundances and element-to-element ratios that are produced in $i$-process nucleosythesis.
\item Population-synthesis models, which incorporate the physics of the detailed models described above, are necessary to study if the properties of the observed CEMP-$r/s$ population can be reproduced within the binary mass-transfer scenario, in particular the frequency of CEMP-$r/s$ stars among CEMP-$s$ stars and their chemical abundances. For this scenario to work, low-mass AGB stars down to approximately $1.2\,\Msun$ have to experience $i$-process nucleosynthesis. Should it be proved that the $i$-process is activated at low masses and metallicities, the binary mass-transfer scenario would be confirmed as a robust and most likely mechanism to explain the formation of CEMP stars enriched in neutron-capture elements. This would also have important consequences for our understanding of the early chemical enrichment of galaxies, because it would support the hypothesis that all CEMP stars enriched in neutron-capture elements have a common origin, which is different from CEMP-no stars, as suggested by many authors \cite[e.g.][]{Spite2013, Starkenburg2014, Bonifacio2015}.
\end{itemize}


\section{Conclusions}
\label{concl}

None of the models as yet proposed to explain the origin of CEMP-$r/s$ stars currently reproduces the properties of the observed CEMP-$r/s$ population. In particular, in all but one formation scenarios the observed frequency of CEMP-$r/s$ stars is underestimated by at least a factor of five and up to two orders of magnitude. The only model that predicts a ratio of CEMP-$r/s$ to CEMP-$s$ stars almost consistent with the observations fails to reproduce the correlation between the abundances of europium and barium observed in CEMP stars, and overestimates the proportion of $r$-rich stars by more than a factor of ten. 

It has been proposed that the \emph{intermediate} or $i$-process may be activated in some circumstances in AGB or post-AGB stars. Preliminary results show that the theoretical abundance distributions predicted by the models are consistent with those observed in CEMP-$r/s$ stars. CEMP-$r/s$ stars could therefore be the secondary stars of binary systems that in the past accreted material from the winds of AGB primary stars, that is, the same formation scenario proposed for CEMP-$s$ stars. Further calculations are necessary to determine at what masses and metallicities the $i$-process is triggered in AGB stars, what abundance distribution is produced by the $i$-process, and how likely it is to form CEMP-$r/s$ stars. If the preliminary results were confirmed, the binary mass-transfer scenario would stand out as a robust explanation of the origin of all CEMP-$s$ stars, including CEMP-$r/s$ stars. This would support the hypothesis that metal-poor stars highly enriched in neutron-capture elements are formed in binary systems, whereas other carbon-normal or carbon-enhanced metal-poor stars have a different formation history.


\begin{acknowledgements}
The authors are grateful to Dr. R. Izzard for his untiring support with the code \texttt{binary\_c}. RJS is the recipient of a Sofja Kovalevskaja Award from the Alexander von Humboldt Foundation. 
\end{acknowledgements}



\end{document}